\documentclass[12pt]{article}
\usepackage{amsthm,graphicx,amssymb}

\theoremstyle{remark}

\newcommand*{\eqref}[1]{(\ref{#1})}

\DeclareSymbolFont{iso}{U}{txmia}{m}{it}
\DeclareMathSymbol{\rea}{\mathalpha}{iso}{"92}
\DeclareMathSymbol{\okr}{\mathalpha}{iso}{"93}
\DeclareMathSymbol{\nab}{\mathalpha}{iso}{"6E}
\DeclareMathSymbol{\isom}{\mathalpha}{iso}{"24}

\date{}
\title {Classes of exact static spheroidal Einstein-Maxwell solutions.}
\author{S.M.KOZYREV\thanks{Scientific center gravity wave studies ''Dulkyn'',
PB 595, Kazan, 420111, Russian Federation. \emph{Email}:
Sergey@tnpko.ru}}
\begin{document}
\maketitle

\begin{abstract}
In this paper we study the spheroidal cases of static charged
fluid configurations in general relativity. We consider the effect
of the anisotropic stresses of electromagnetic field on the shape
of static charged self-graviting objects. It is shown that
electromagnetic fields can have significant effect on the
structure and properties of self-graviting objects.
\end{abstract}

\section{Introduction}
Exact solutions of the Einstein-Maxwell field equations are of
crucial importance in relativistic astrophysics. These solutions
may be utilised to model a charged relativistic star as they are
matchable to the Reissner-Nordstrom \cite{Reissner},
\cite{Nordstrom} exterior at the boundary.
 A
wide spread assumption in the study of stellar structure is that
the shape of star can be modeled as a spherical symmetry object.
This approach has been used extensively in the study of star, star
system and galaxies. However, in many systems, deviation from
spherical symmetry may play an important role in determining of
them properties. Physical situation where unspherical shape may be
relevant are very diverse. On the other hand, self-graviting
objects resulting from the coupling electromagnetic field to
gravity are a system where anisotropic pressure occurs naturally.

 Anisotropy appears as an extra assumption on the
behavior of electromagnetic field and on the shape of equilibrium
configuration. Since we still do not have a formulation of the
possible anisotropic stresses is emerging in these or other
contexts, we take the approach of finding several exact solutions
representing physical situations, modelled by ellipsoid of
revolution. Solutions to the equation in spheroidal coordinates
have application to a wide range of problems in physics \cite
{Flammer}. Our goals hear is to find exact spheroidal solution,
offering an analysis of the change in the physical properties of
the stellar and galaxy models due to presence of electromagnetic
field.

\section{Einstein-Maxwell Equations and Static spheroidal configurations.}
In this paper we study static, spheroidal solutions of the
Einstein-Maxwell system featuring a spinless charge
configurations.The vacuum Einstein-Maxwell equations, in
geometrized units such that $c = 8\pi G = \mu_0 = \varepsilon_0 =
1$, can be written as \cite{Tolman}
\begin{equation}
R_{\mu \nu }-\frac 12Rg_{\mu \nu }= T_{\mu \nu }, \label{eqf1}
\end{equation}
\begin{equation}
F^{\mu \nu }  _{; \nu} = 0,  \label{eq6}
\end{equation}
with the electromagnetic energy-mementum tensor given by
\begin{eqnarray}
T_{\mu \nu } &=& F_{\mu \eta}F^{ \eta}_{\nu }- \frac 14 g_{\mu \nu
}F_{ \eta \zeta}F^{\eta \zeta }, \label{eq5}
\end{eqnarray}
where
\begin{eqnarray}
F_{\mu \nu} &=&A_{\nu,\mu }-A_{\mu ,\nu}, \label{eq3}
\end{eqnarray}
is the electromagnetic field tensor and $A_\mu $ is the
electromagnetic four potential.

 To start with, note that by using coordinate freedom
inherent in general relativity any static spheroidal geometry can
by put into form where are only two independent metric components
typically functions of the coordinates $\xi$. As we have already
mentioned we consider the ansatz static spheroidal space-time.

 The two-dimensional elliptic coordinate
system is defined from the set of all ellipses and all hyperbolas
with a common set of two focal points. We denote the separation of
the two focal points by $2c$.

 Oblate
spheroidal coordinates are derived from elliptic coordinates by
rotating the elliptical coordinate system about the perpendicular
bisector of the focal points. The coordinates are often labelled
$\eta$, $\xi$ and $\theta$ with the transformation to Cartesian
coordinates given by

\begin{eqnarray}
x =c\eta \xi sin(\theta), \label{eqXo}
\end{eqnarray}
\begin{eqnarray}
y =c \sqrt{\left(\xi^2-1\right) \left(1-\eta^2\right) },
\label{eqYo}
\end{eqnarray}
\begin{eqnarray}
z =c\eta \xi sin(\theta)cos(\theta). \label{eqZo}
\end{eqnarray}

Similarly, one can obtain the prolate spheroidal coordinates by
rotating it about the parallel bisector.

\begin{eqnarray}
x =c\eta \xi, \label{eqX}
\end{eqnarray}
\begin{eqnarray}
y =c \sqrt{\left(\xi^2-1\right) \left(1-\eta^2\right)
}cos(\theta), \label{eqY}
\end{eqnarray}
\begin{eqnarray}
z =c \sqrt{\left(\xi^2-1\right) \left(1-\eta^2\right)}sin(\theta).
\label{eqZ}
\end{eqnarray}

Let the spacetime ansatz be described by the spheroidal metric
given by
\begin{eqnarray}
ds^2=-B \left( \xi\right) dt^2+  A \left( \xi\right) d\Omega^2.
\label{eqSph}
\end{eqnarray}
For this static spheroidal  ansatz of space-time we take the
electromagnetic potential as
\begin{eqnarray}
A_\mu &=&(\psi,0,0,0), \label{eqEp}
\end{eqnarray}
where it is assumed that the electric potential $\psi$ depends on
$\xi$ only.

 We adopt coordinates that allow us to
write spheroidal geometry in prolate form
\begin{eqnarray}
d\Omega^2= c^2
 \frac{\xi^2-\eta^2}{\xi^2-1 }  d \xi ^2+c^2
 \frac{\xi^2-\eta^2}{1 - \eta^2}  d\eta^2+c^2
 (\xi^2-1)(1-\eta^2) d\theta^2,
\label{eqSphPr}
\end{eqnarray}
and in oblate form
\begin{eqnarray}
d\Omega^2= c^2 \frac{\xi^2-\eta^2}{\xi^2-1 }  d \xi ^2+c^2
\frac{\xi^2-\eta^2}{1 - \eta^2}  d\eta^2+(c
 \xi \eta)^2 d\theta^2,
\label{eqSphOb}
\end{eqnarray}
where $A, B$ are function of $\xi$ only and $\xi \geq  1$,  $-1
\leq \eta \leq 1$, $0 \leq \theta \leq 2\pi$.

\subsection{Prolate spheroidal configurations.}

After a bit of algebra, the  field equations (\ref{eqf1}) -
(\ref{eq6}) are explicitly given in forms of the metric
(\ref{eqSph}) in prolate case.

\begin{eqnarray}
\frac{A'^2}{4A^2}+\frac{A'B'}{2A B}
=-\frac{\varepsilon^2}{2(1-\xi^2)A}=T_{11}, \label{eq4p1}
\end{eqnarray}

\begin{eqnarray}
\frac{\xi^2-1}{2(\eta^2-1)}\left[-\frac{A''}{A}
-\frac{B''}{B}+\frac{A'^2}{A^2}+\frac{B'^2}{2B^2}
 \right]=\frac{\varepsilon^2}{2(\xi^2-1)(\eta^2-1)A}=T_{22}, \label{eq4p2}
\end{eqnarray}

\begin{eqnarray}
 \frac{A'}{A}+\frac{B'}{B} =0=T_{12}, \label{eq4p21}
\end{eqnarray}

\begin{eqnarray}
\frac{(\xi^2-1)(\eta^2-1)}{2(\xi^2-\eta^2)}\left[-\frac{A''}{A}
-\frac{B''}{B}+\frac{A'^2}{A^2}+\frac{B'^2}{2B^2} \right]
=-\frac{\varepsilon^2(\eta^2-1)}{2(\xi^2-\eta^2)A} =T_{33},
\label{eq4p3}
\end{eqnarray}

\begin{eqnarray}
\frac{B(\xi^2-1)}{c(\xi^2-\eta^2)}\left[-\frac{A''}{A^2} +
\frac{3A'^2}{4A^3}- \frac{2\xi}{\xi^2-1} \frac{A'}{A^2}\right]
=\frac{\varepsilon^2 B}{2
c(\xi^2-1)(\xi^2-\eta^2)A^2}=T_{00},\label{eq4p4}
\end{eqnarray}
where prime (') denoting derivative with respect to the $\xi$
coordinate.

 After a simple integration, from (\ref{eq4p1}) - (\ref{eq4p4}) we obtain

\begin{equation}
A=\frac
18\left[a_0\pm\varepsilon\ln\left(\frac{\xi+1}{\xi-1}\right)
\right]^2, \nonumber
\end{equation}
\begin{equation}
B =\frac{b_0}  {A }, \label{e5}\\
\end{equation}
where  $a_0, b_0$ arbitrary constants.

\subsection{Oblate spheroidal configurations.}
Replacing the line element in the field equations, the oblate set
is
\begin{eqnarray}
 \frac{A'^2}{4A^2}+\frac{A'B'}{2A
B}=-\frac{\varepsilon^2 }{2\xi^2(\xi^2-1)A}=T_{11}, \label{eq4f1}
\end{eqnarray}

\begin{eqnarray}
\frac{\xi^2-1}{2(\eta^2-1)}\left[- \frac{A''}{A} -\frac{B''}{B} +
\frac{A'^2}{A^2}+\frac{B'^2}{2B^2} \right]=-\frac{\varepsilon^2
}{2\xi^2(\eta^2-1)A}=T_{22}, \label{eq4f2}
\end{eqnarray}

\begin{eqnarray}
\frac{A'}{A}+\frac{B'}{B} =0=T_{12}, \label{eq4f21}
\end{eqnarray}

\begin{eqnarray}
\frac{\xi^2\eta^2(\xi^2-1)}{2(\xi^2-\eta^2)}\left[\frac{A''}{A}
+\frac{B''}{B}-
\frac{A'^2}{A^2}-\frac{B'^2}{2B^2}\right] \nonumber \\[0.01in]
=-\frac{\varepsilon^2\eta^2}{2(\xi^2-\eta^2)A}=T_{33},
\label{eq4f3}
\end{eqnarray}

\begin{eqnarray}
\frac{B(\xi^2-1)}{c(\xi^2-\eta^2)}\left[-\frac{A''}{A^2} +
\frac{3A'^2}{4A^3}- \frac{(2\xi^2-1)}{\xi(\xi^2-1)}
\frac{A'}{A^2}\right]
=\frac{B\varepsilon^2}{2c\xi^2(\xi^2-\eta^2)A^2}=T_{00},
\label{eq4f4}
\end{eqnarray}

 Then the solutions of the gravitational field
equations take in oblate case the form
\begin{equation}
A=2 \left(a_0\pm\varepsilon \arctan\sqrt{\frac{\xi^2-1}{\xi^2+1}}
\right)^2, \nonumber
\end{equation}
\begin{equation}
B =\frac{ b_0}{A }, \label{e4}\\
\end{equation}

where $ a_0, b_0$ arbitrary constants.

\subsection{Analysis }

 To see that all these metrics is
asymptotically flat Minkowski it is enough to show that the metric
components behave in an appropriate way at large $\xi$-coordinate
values, e.g., $g_{\mu \nu } = \eta_{\mu \nu }+ O(1/\xi)$ as $\xi
\rightarrow \infty $. By inspection of the coefficients, we verify
that this is so ($a _0=1, b_0 =1$).

In fact, in the present approach, it is easy to show that, in the
case of absent the electromagnetic field $\varepsilon=0$,
Einstein's field equations yield only the flat space

\begin{equation}
A=1, \nonumber
\end{equation}
\begin{equation}
B =1, \label{e5en}\\
\end{equation}
\begin{equation}
\phi =1. \nonumber
\end{equation}

Therefore, we see that it is possible to explain the shape of
spheroidal configurations by electromagnetic or other fields
\cite{SM}. This seems to be a remarkable result, although in a way
it should be anticipated since the directional components of
"equation of state" of electromagnetic field are anisotropic in
the oblate and prolate cases. However, in this case there is a
contribution from the electromagnetic field that makes $T_{\mu \nu
}$ nonzero. On the other hand, it seems natural that we have
obtained an "equation of state" that describes vacuum, since we do
not have matter.

\section{Discussion}
In this article we delineated the qualitative features one would
expect from spheroidal object. It is demonstrated that our model
can successfully predict the spheroidal configuration in terms of
a self-gravitating spacetime solution to the Einstein field
equations and reproduce the not spherically-symmetric shape in
terms of the non-trivial energy density and anisotropic pressure
of the electromagnetic field which was absent in the context of
empty space.

Hence the approach followed in this paper has proved to be a
fruitful avenue for generating new exact solutions for describing
the spacetimes of charged configurations.

We believe that following this hypotheses the shape of galaxy and
rotation curve may be explained by action of electromagnetic or
other fields. The solution presented here could be a first
approximation at the galactic space-time provided the presence of
any physical fields. Therefore, it is necessary to study how these
results modify the standard method of interpretation rotation
data. Further investigation into the nature solutions with view to
separating the real rotational effects from the electromagnetic,
scalar or other fields anisotropy might be rewarding.

\section{Acknowledgements} \nonumber
I am grateful to S.V. Sushkov and R.A. Daishev  for the helpful
discussions. The work was supported in part by the Institute of
Applied Problems.

\end{document}